# Numerical calculation of near field scalar diffraction using angular spectrum of plane waves theory and FFT

**Adrián Carbajal-Domínguez[1], Jorge Bernal Arroyo[1], Jesús E. Gómez Correa[1], Gabriel Martínez Niconoff[2]**

(1) Universidad Juárez Autónoma de Tabasco, División Académica de Ciencias Básicas, Cunduacán, Tabasco, México, C.P. 86690,

adrian.carbajal@dacb.ujat.mx

(2) Instituto Nacional de Astrofísica, Óptica y Electrónica (INAOE), Departamento de Óptica, Grupo de Óptica Estadística, Apdo. 51 y 216, Puebla, México, C. P. 72000

*Abstract*

It is a known fact that near field diffraction or Fresnel diffraction calculations are difficult to perform exactly. It is in general necessary to make some approximations in order to obtain a more suitable form. In this work, a numerical implementation based on angular spectrum theory for near field diffraction is presented. The method uses Fast Fourier Transforms (FFT) and it turns out to be accurate and fast. In order to show the capabilities of the method, diffraction near field for a circular aperture and a spiral slit are studied. Numerical and experimental results are shown. This method could be useful to implement pc based physical optics learning.

# 1. - Introduction

As an optics student and as teacher, one can easily realize that near field diffraction integrals are very difficult to perform, even when numerical methods are used. From a mathematical point of view, scalar wave diffraction in optics is a partial differential equation, known as Helmholtz equation and a boundary condition or transmittance function. If the observation plane is rather close to the transmittance function under free propagation, then Fresnel diffraction appears and as it occurs in most cases, it cannot be calculated without introducing some kind of approximations for phase terms. Roughly, there are two descriptions that allow solving this problem. One is the Kirchhoff-Fresnel theory in which by using Green theorem an exact solution can be obtained by adding two Helmholtz equation solutions. One of both represents transmittance function and the other represents the optical field, usually, it is free space Green function $\exp(ikr)/r$. These boundary conditions are over specified and are mathematically inconsistent [1] because diffracted field and its derivative have to be zero in some part of the integration trajectory. One way to solve such problem is by using Rayleigh-Sommerfeld diffraction theory in which derivative function on the boundary has not to be specified [2] because a new kind of Green functions that vanish at transmittance plane are introduced.

A second major problem arises when this diffraction theory is applied to plane transmittance functions in near field diffraction. In this case, as the distance between transmittance function and observation plane tends towards zero, high frequency oscillations appear which make impossible to evaluate Huygens-Fresnel diffraction integral because it has a pole at the origin and neither Fresnel approximation is fulfilled. Some authors have treated this problem by introducing some kind of approximations mainly for quadratic phase terms [3]. Nevertheless, these solutions continue to have problems when evaluated for distances close to the transmittance function.

An alternative theory is known as angular spectrum of plane waves (ASPW) [2,4], in this case boundary condition, which is the transmittance function, has to be Fourier transformed and it gives the associated amplitudes for plane waves traveling in all possible directions in space. By construction, the solution is well defined at the origin. This means that diffracted field can be calculated at arbitrary close observation distances.

For practical reasons, it is preferred to perform a numerical simulation before the experiment. In diffraction theory, numerical methods based on convolution [2], quadratic phase terms approximations [3] or in terms of Lommel functions expansions are known [5]. Although they correctly predict diffraction for some regions, they all fail at regions near to transmittance function. Recently [6], it has been reported a numerical method for Rayleigh-Sommerfeld diffraction integral as compared to ASPW. Nevertheless, only homogeneus modes are considered which causes some differences with exact results [7]. The effects of this are a) solutions do not fulfill Sommerfeld radiation condition, *i.e.* diffracted field never decays as $z \to \infty$; b) edge effects are excluded. Besides, ASPW recovers Rayleigh diffraction intergal by using Weyl representation for spherical waves [4].

In present work a simple numerical method based on ASPW is introduced in order to perform near field diffraction calculation. The method uses two Fast Fourier Transforms (FFT), which are relatively easy to implement in most programming environments given that many FFT routines and source codes are freely available [8,9]. Commercially available software might also be a solution for less skilled programmers because most of them include their own FFT implementations in 1D or 2D. In order to illustrate this interesting feature two cases are studied. First, near field diffraction for a circular aperture is obtained for arbitrary small observation distances in which other known to the author methods fail. As a second example near field diffraction for a single arm Euclidean spiral slit is presented. A third example is a project in which digital Fresnel Holography is studied. In our, experience, this technique has proven useful to teach diffraction theory to undergraduate physics students exploring more complex projects.

## 2. - Angular spectrum theory

The objective is to find exact solutions to Helmholtz equation

$$\nabla^2 \phi + k^2 \phi = 0 \tag{1}$$

with boundary condition at transmittance function position $z = 0$ given by

$$\phi(x, y, z = 0) = t(x, y) \tag{2}$$

In particular, one asks for a solution written as an infinite and continuous superposition of plane waves

$$\phi(\vec{r}) = \int_{-\infty}^{\infty} A(\vec{k}) \exp(i\vec{k} \cdot \vec{r}) d\vec{k} \tag{3}$$

where $\vec{r} = x\hat{i} + y\hat{j} + z\hat{k}$ is the vector position and

$$\vec{k} = \frac{2\pi}{\lambda} \hat{e}_k \tag{4}$$

the wave vector. The function $A(\vec{k})$ defines the amplitude for each plane wave in each direction given by unit vector $\hat{e}_k = \cos\alpha \hat{i} + \cos\beta \hat{j} + \cos\gamma \hat{k}$. In order to simplify notation, parameters are defined

$$u = \frac{\cos\alpha}{\lambda}; v = \frac{\cos\beta}{\lambda}; p = \frac{\cos\gamma}{\lambda} \tag{5}$$

and the geometry is shown in fig (1). In this case, the dot product takes the form

$\vec{k} \cdot \vec{r} = 2\pi(xu + yv + zp)$ and eq. (3) can be written as

$$\phi(x,y,z) = \int_{-\infty}^{\infty}\int_{-\infty}^{\infty}\int_{-\infty}^{\infty} A(u,v,p)\exp[i2\pi(xu + yv + zp)]dudvdp \qquad (6).$$

Using eq.(6) as a solution to eq. (1) leads to the condition

$$\frac{1}{\lambda^2} = (u^2 + v^2 + p^2) \qquad (7)$$

This expression allows us to reduce eq. (6) to only to integrals

$$\phi(x,y,z) = \int_{-\infty}^{\infty}\int_{-\infty}^{\infty} A(u,v)\exp[i2\pi(xu + yv + zp)]dudv \qquad (8)$$

Eq. (7) defines a sphere in the $u, v, p$ space, as shown in fig (1) insert, and one can express $p = p(u,v)$ in the form

$$p(u,v) = \begin{cases} \sqrt{\dfrac{1}{\lambda^2} - u^2 - v^2} & \text{if } \dfrac{1}{\lambda^2} > u^2 + v^2 \\ i\sqrt{u^2 + v^2 - \dfrac{1}{\lambda^2}} & \text{if } u^2 + v^2 > \dfrac{1}{\lambda^2} \end{cases} \qquad (9)(a,b)$$

Eq. (9a) gives rise to homogeneous or propagating waves, which have a major contribution to the far field whilst eq. (9b) represents evanescent waves. Applying the boundary condition defined by (2) to eq. (8), leads us to the angular spectrum

$$A(u,v) = \int_{-\infty}^{\infty}\int_{-\infty}^{\infty} t(x,y)\exp[i2\pi(xu + yv)]dxdy \qquad (10)$$

which gives the amplitudes for every propagation direction. Rewriting eq.(8) in the form

$$\phi(x,y,z) = \int_{-\infty}^{\infty}\int_{-\infty}^{\infty} A(u,v)\exp[i2\pi zp(u,v)]\exp[i2\pi(xu + yv)]dudv \qquad (11)$$

it allows us to recognize Fourier transform for the convolution between $t(x,y)$ function and Fourier transform of $\exp[i2\pi zp(u,v)]$, which behaves as a propagation function because it only depends on coordinate $z$ and it also defines diffracted field as a summation of modes. At $z = 0$, eq. (11) remains defined and recovers transmittance function as expected. Any value around $z = 0$ is also defined and permits the evaluation of eq.(11) for arbitrary small observations distances. It is also remarkable that we can obtain the diffracted field after only two Fourier transforms. Other methods based on Huygens-Fresnel integral require about 3 Fourier transforms [2]. Symmetry based methods [10] that calculate spherical waves emerging from every point in transmittance function have to deal with an enormous number of distances calculations. In this case, distance calculations performed in other methods are replaced by the simpler $z$ coordinate, which represents the distance form the origin to the observation plane. .

Eq.(9) implies that two kinds of contributions are possible. Propagating modes as in $p(u,v) = \sqrt{1/\lambda^2 - u^2 - v^2}$ and evanescent ones when $p(u,v) = i\sqrt{u^2 + v^2 - 1/\lambda^2}$. Both cases depend on how $1/\lambda^2$ is as compared to $u^2 + v^2$ term. In traditional Fresnel diffraction theory is assumed that transmittance function dimensions are much bigger than wavelength because edge effects are neglected. In the ASPW case, edge effects are at the high frequencies of the Fourier transformed transmittance function, as can be seen in eq.(10). In other words, edge effects appear in the $\lim_{u,v \to \infty} A(u,v)$. This can be better understood if we consider writing eq. (11) in the form

$$\phi(x,y,z) = \int\int_{\frac{1}{\lambda^2} > u^2+v^2} A(u,v)\exp[i2\pi z p(u,v)]\exp[i2\pi(xu+yv)]dudv$$
$$+ \int\int_{\frac{1}{\lambda^2} < u^2+v^2} A(u,v)\exp[-2\pi z p(u,v)]\exp[i2\pi(xu+yv)]dudv \quad (12)$$

in which the first integral term represents propagating modes and the second corresponds to evanescent ones. In order to find physically valid solutions, both contributions most be considered. If only propagating solutions are accounted for, then one would obtain diffracted fields propagating until $z$ with very large values and Sommerfeld radiation condition would be violated. In the other hand, when both contributions are considered, then one would obtain fields with decreasing intensity as $z$ becomes large.

In the other hand, it can be shown [4] that in the far field, ASPW recovers Fraunhofer diffraction as expected as shown in eq.(12).

$$\phi(x,y,z) \approx (-2\pi i/k)(z/r)A(x/r,y/r)\exp[ikr]/r \qquad (13)$$

So far, we have seen that ASPW provides to us an exact solution to Helmhlotz equation; it also allows the calculation of diffracted field at arbitrarily small distances; edge effects are considered; the solution comprises evanescent and propagating contributions; and at the far field it recovers Fraunhofer diffraction.

Another problem arises when specific problems need to be studied. For instance, let us consider the diffraction from an annular slit of radius D, given by a Dirac Delta function in polar coordinates [11]

$$t(r,\theta) = \frac{\delta(r-D)}{r\pi} \qquad (14)$$

In this case, after writing equations (8) and (10) in cylindrical coordinates, we get the spectrum $A(\rho) = 2J_0(2\pi D\rho)$ and the near field is of the form

$$\phi(r,z) = 4\pi \int_0^\infty J_0(2\pi D\rho)J_0(2\pi r\rho)\exp[i2\pi z(1/\lambda^2 - \rho^2)^{1/2}]\rho d\rho \qquad (15)$$

Eq. (15) is very difficult to integrate analytically and numerically. In the first case, the exponential term has a radical in its argument which makes it very hard to integrate if no approximation is made. In the numerical case, Bessel functions need to be expressed as infinite series expansions or need to be written by using polynomials approximations

[12]. Either way, numerical calculation becomes very complex and computationally slow, because of numerical integration and/ or series convergence, which in this case is the worst because the number of terms considered in the Bessel series has to be much bigger than the Bessel function argument. One way to overcome the problem is to device a method of calculation that employs FFT algorithms in order to increase computational speed.

### 3.-Numerical implementation based on 2D FFT

For the numerical solution, it is necessary to transcribe the problem in discrete form. Transmittance function is now represented by means of a $n \times m$ matrix in the form or a bitmap image

$$T_{m,n} = t(m,n), \qquad (16)$$

where *m*, *n* are integers and we also consider that in such matrix there is an object (a slit, for instance) of "radius" or size $R$, in pixels. In practical terms, transmittance $T_{m,n}$ is an image file, in this case formed by $n \times m$ pixels. Then standard 2D fast Fourier transform (FFT) is calculated and the resulting matrix is re-centered to correct low and high frequencies misplacement that is a consequence of discrete FFT algorithm. In this way, one obtains $A(u,v)$ in discrete form or

$$A^{discrete} = FFT\{T\}. \qquad (17)$$

Then, a *G* matrix is defined as

$$G(z)_{m,n} = \exp[i2\pi z p(m,n)]. \qquad (18)$$

Equation (9) of function $p(x,y)$ defines propagation direction for all the superposed plane waves. In present case, we have noticed that values of the form $\lambda = 1/(Rq)$ where $R$ is the object radius estimated in pixels, $q \geq 2$, $z < 10$ work well to obtain near field diffraction because in this way evanescent and propagating contributions are reasonably included. In order to simplify calculations we consider squared images of $200 \times 200$ pixels, but increasing the size would increase computer time proportionally to a couple FFT calculations of such size.

The product element by element between eq. (17) and eq. (18) is made, leading us to

$$H_{m,n} = A_{m,n} G(z)_{m,n}, \qquad (19)$$

matrix *H* being an intermediate step. Then, inverse fast Fourier transform is performed,

$$\Phi = FFT^{-1}(H), \qquad (20)$$

which give us $\Phi$ the matrix representing calculated diffracted field. Finally, a logarithmic transformation for pixel intensities is made in order to obtain a clearer image. A flowchart of this algorithm is shown in fig. (3). It is worth noticing that this method needs only two FFT operations. Moreover, no approximation at all has been made (apart of those inherent to FFT algorithm and the discrete representation of the field) which implies that the present method is accurate and faster than standard methods [Goodman] based on convolution theorem for Huygens-Fresnel diffraction in which three FFT are required and phase approximations are made.

## 4. - Examples

### *4.1 Circular aperture*

In fig. (3) numerical implementation of eq.(4) is shown for different and $z$ values for a circular aperture. In fig. (3a) appears circular transmittance function. In fig. [3 (b-c)] diffraction patterns are shown for different propagation values and characteristic features as Fresnel Zones and Poisson spot can be observed. The method is relatively simple to program and to evaluate for a given distance $z$ and it is natural to implement a program that iteratively calculates the diffracted field for an interval of $z$ distances starting at zero. In each distance only the axial intensities are retained and all these vectors are stacked forming an image of diffracted field as if viewed perpendicularly to the propagation direction, as shown in fig. (4). In fig. (5) an axial intensity plot of results obtained in fig. (4) is presented. It is clear that diffracted field can be obtained even for distances very near to the aperture without undesired high frequency oscillations. Axial intensity distribution is calculated in order to compare to other reported methods [3,13] and it is shown in fig.(5) insert.

## *4.2 Spiral slit*

When transmittance function is a slit, focusing regions are related to geometry, in particular to evolutes which are centers of curvature [14]. A spiral slit has interesting curvature features. Besides, diffracted field for spiral slits has been studied [15] only for the far field case. Euclidean single arm spiral is considered as shown in fig. (6a). The corresponding calculated diffracted fields for different $z$ values fig.[6(b,c)] are shown and the experimental results are shown in fig.[6(d,e)]. It is observed in simulation and in experiment that lobular structures appear. The leaf number decreases as $z$ coordinate augments. Detailed study for spiral slits will be presented elsewhere. It is remarkable how accurate this numerical calculation is as compared to experiment. Further study is needed to understand all features of these fields.

## *4.3 Other projects*

This technique, in our personal teaching experience, has proven useful to study other physical optics phenomena. One of our interest is generation of non diffracting beams, particularly, $J_0$ Bessel beams [16] which are easily produced by considering an annular slit. Profiles intensity for different propagation distance can be plotted rapidly with very little effort. Besides, a projection of diffracted field, viewed perpendicularly to propagation direction can be plotted by calculating a number of diffracted fields for different propagation distance values, then, only axial transversal intensities are retained and stacked to form an image. An example for a circular aperture is shown in fig.

Computer generated Fresnel Holograms [17] can be studied recursively in this context as shown in fig. (7). First, one has to calculate diffracted field to a fixed propagation distance added to a plane wave. The resulting field, the hologram, is saved. Then, this hologram can be numerically reconstructed by using it as a transmittance function. In the other hand, if the calculated hologram is photographed, it can be reconstructed in the laboratory by using a plane wave form any laser source.

## 5. - Conclusion

A numerical method for evaluating near field diffraction has been introduced by using angular spectrum of plane waves. The major advantage is that high frequency phase oscillations near transmittance function have been removed and the solution only depends in coordinate propagation $z$. This method can be very useful in areas as computer generated holograms, diffraction free-beams, phase singularity optics or for educational purposes as an interactive and accurate tool for studying scalar diffraction phenomena.

# 6.- References

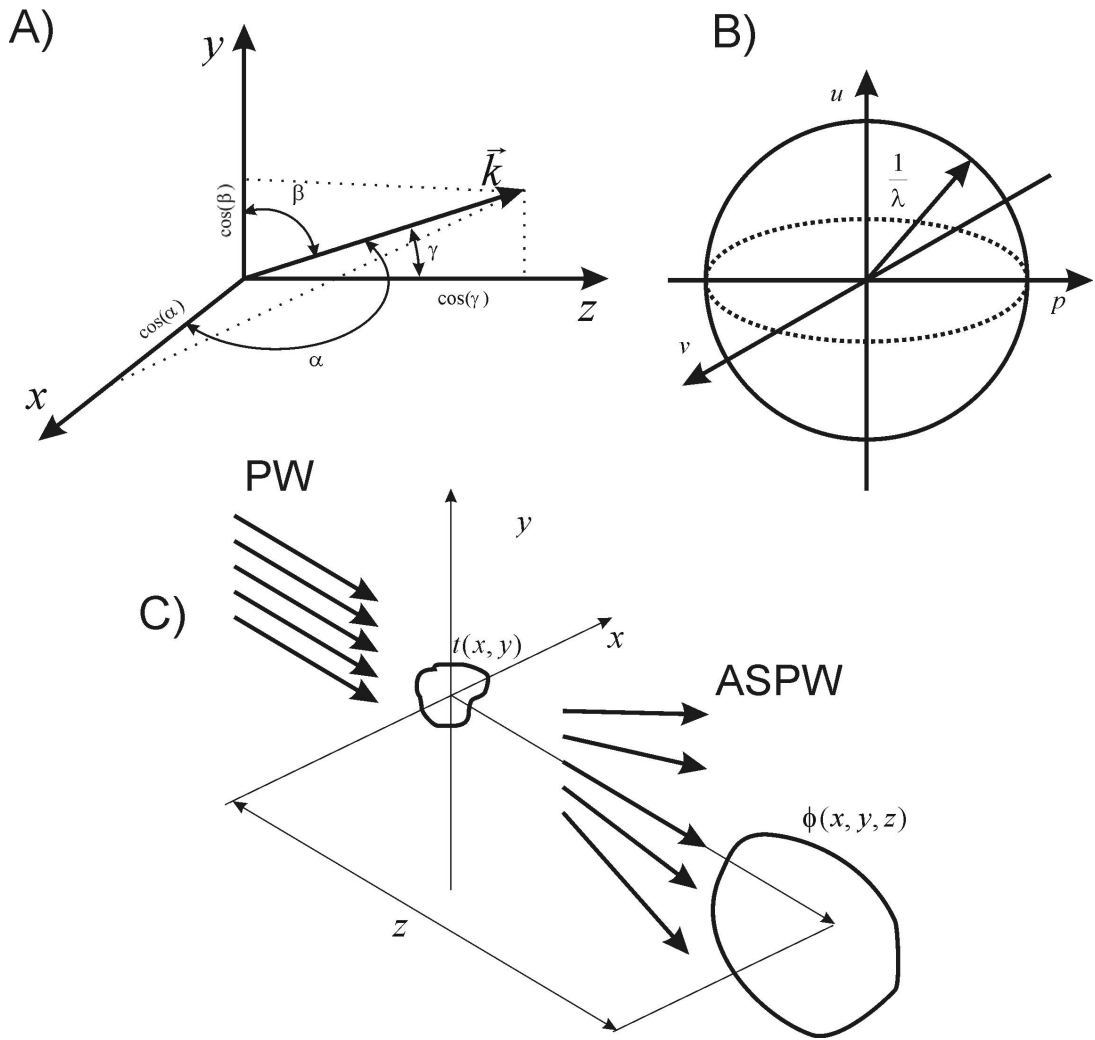

**figure 1** Coordinate system and notation. A) $\alpha,\beta,\gamma$ angles definitions in the $x,y,z$ reference frame. B) Spherical geometry of $u,v,p$ parameters. C) transmittance function $t(x,y)$ and diffracted field $\phi(x,y,z)$ in the $x,y,z$ space. PW incident plane wave; ASPW angulas spectrum of plane waves.

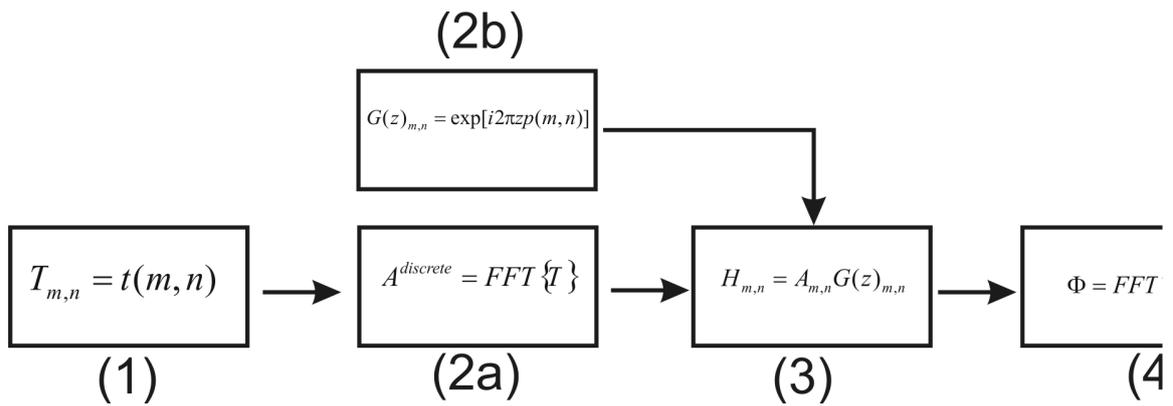

**figure 2** Flow chart showing how to implement the algorithm. For a single calculation, solid lines show the involved steps while for an iterative calculation dotted lines show corresponding flow. For the first case only two FFT are calculated. In the second case, N+1 FFt are calculated, where N is the total number of different z values.

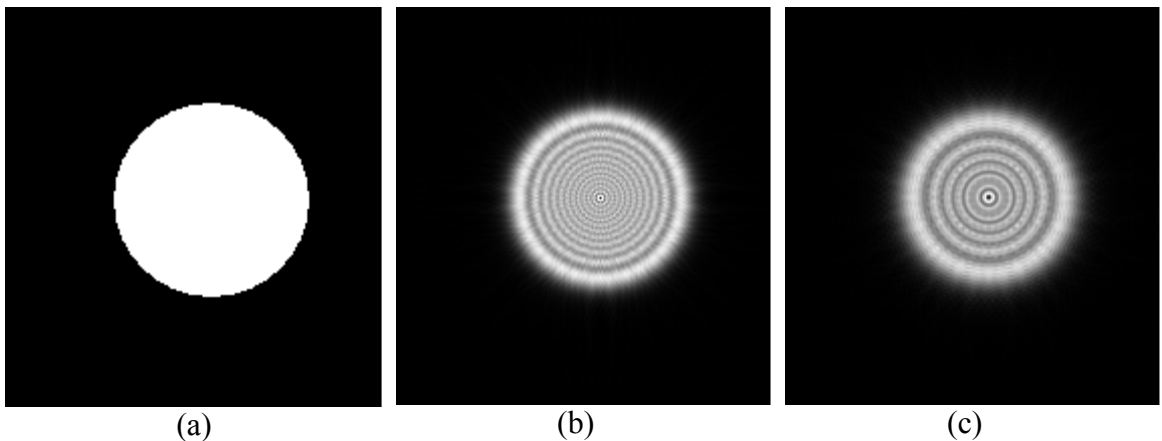

(a) (b) (c)

**figure 3** Numerical simulation for a circular aperture. a) binary image representing transmittance function. b),c) **calculated diffracted field for two different distances. The method reproduces Fresnel zones and Poisson spot.**

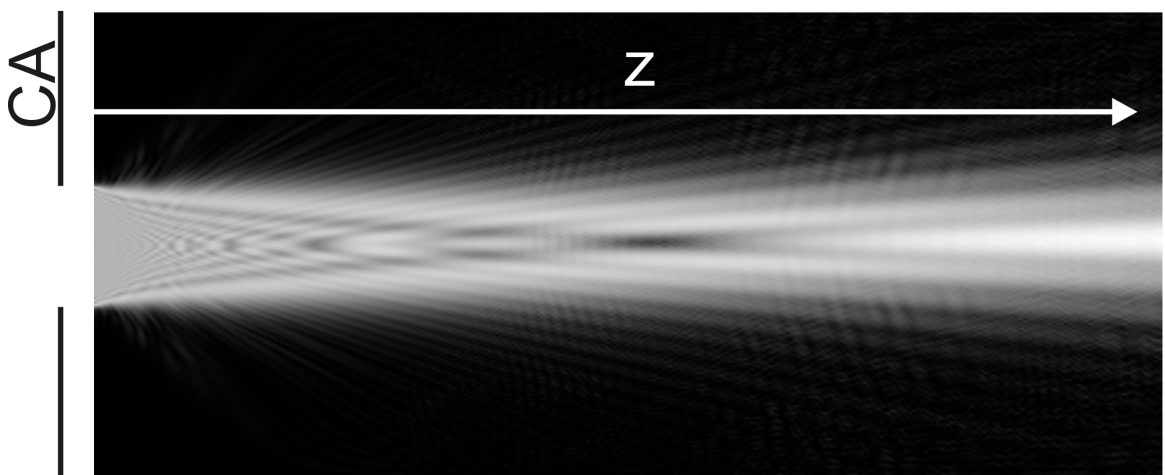

**figure 4** Propagation for different values of z. An ensemble of diffraction patterns are calculated for different values of z. At z =0 the field is defined. CA represents circular aperture position.

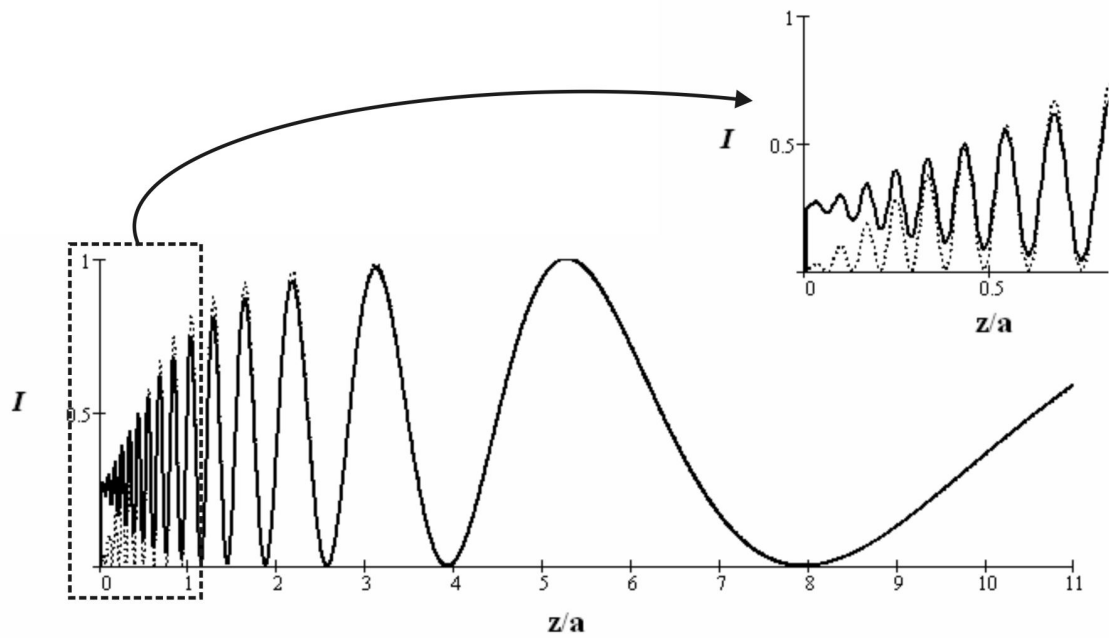

**figure 5**. Axial intensity distribution obtained by numerical integration of eq. (11) (solid) as compared to results reported in ref. [3] (dot) for a circular aperture of radius a.

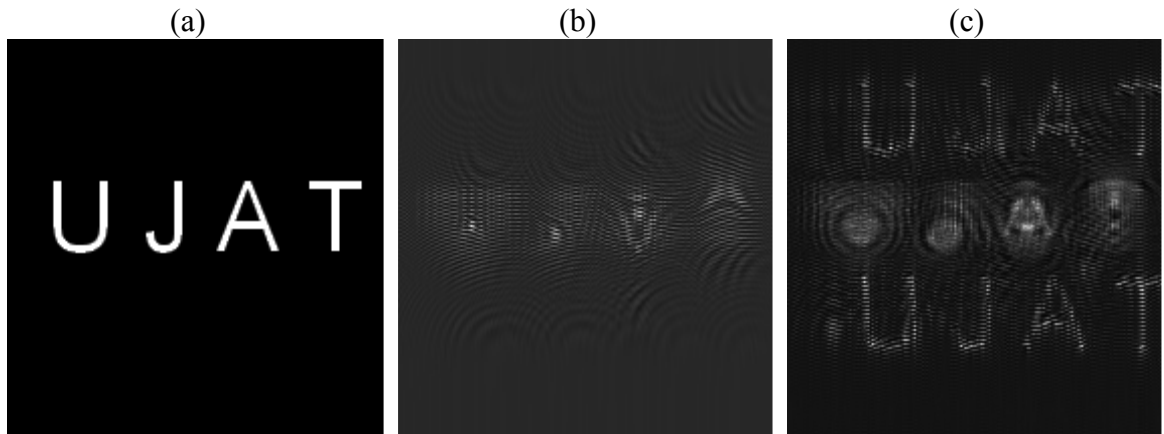

**figure 6** Example of students digital holography project. (a) object, (b) hologram (see text) and (c) reconstruction. Hologram recording and reconstruction are made by using this method.

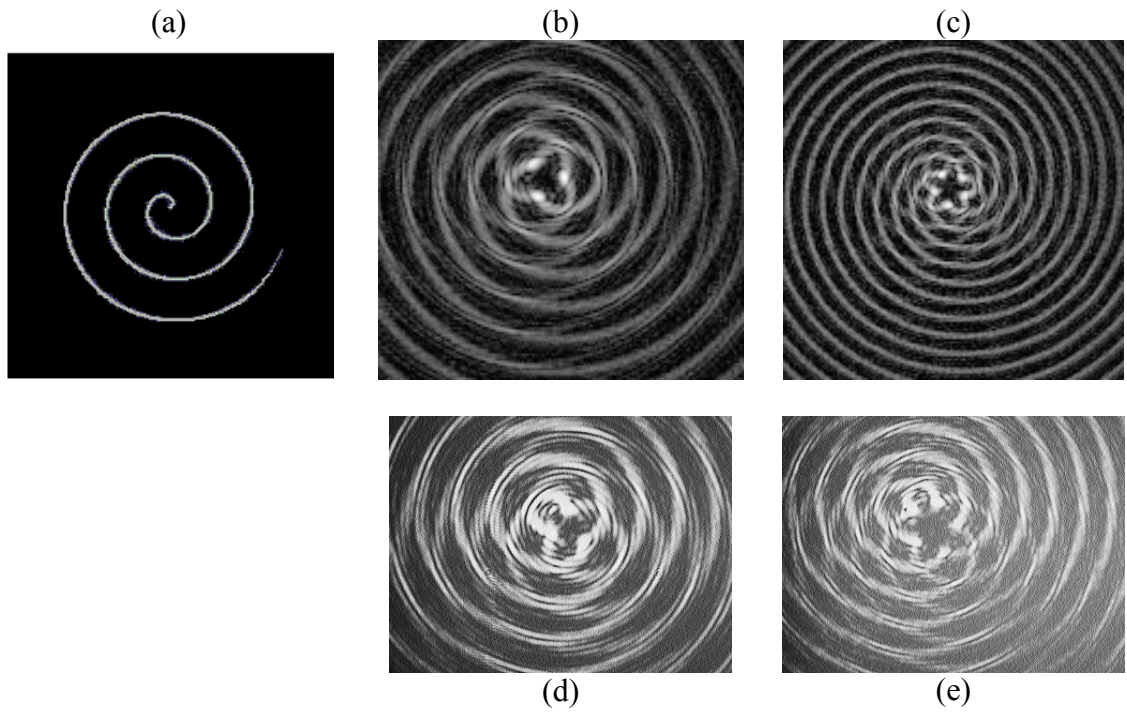
**figure 7** Spiral slit diffraction. (a) Euclidean spiral slit; (b,c) numerical results for two different *z* values. Characteristic 3 and 5 lobes are predicted. (d,e) corresponding experimental results.